\documentclass[prl,twocolumn,showpacs,preprintnumbers,amsmath,amssymb]{revtex4}
\usepackage{amsmath,amssymb}
\usepackage{graphicx}% Include figure files
\usepackage{dcolumn}% Align table columns on decimal point
\usepackage{bm}% bold math

\begin{document}
\title{Frequency Light Shifts Caused by the Effects of Quantization of
Atomic Motion in an Optical Lattice}
\author{A. V. Taichenachev and V. I. Yudin}
\affiliation{Institute of Laser Physics SB RAS, Novosibirsk
630090, Russia} \affiliation{Novosibirsk State University,
Novosibirsk 630090, Russia}
\author{V. D. Ovsiannikov}
\affiliation{Physics Department, Voronezh State University,
Voronezh 394006, Russia}
\author{V. G. Pal'chikov}
\affiliation{Institute of Metrology for Time and Space at National
Research Institute for Physical--Technical and Radiotechnical
Measurements, Mendeleevo, Moscow Region, 141579 Russia}

\date{\today}

\begin{abstract}
Frequency light shifts resulting from the localization effects and
effects of the quantization of translational atomic motion in an
optical lattice is studied for a forbidden optical transition
$J$=0$\to$$J$=0. In the Lamb-Dicke regime this shift is
proportional to the square root from the lattice field intensity.
With allowance made for magneto-dipole and quadrupole transitions,
the shift does not vanish at the magic wavelength, at which the
linear in intensity shift is absent. Preliminary estimates show
that this shift can have a principal significance for the
lattice-based atomic clocks with accuracy of order of
10$^{-16}$-10$^{-18}$. Apart from this, we find that the numerical
value of the magic frequency depends on the concrete configuration
of the lattice field and it can vary within the limits 1-100 MHz
(depending on element) as one passes from one field configuration
to another. Thus, theoretical and experimental investigations of
contributions originated from magneto-dipole and quadrupole
transitions are of principal self-dependent interest.
\end{abstract}


\maketitle

The last few years were marked by a principal ideological \cite{Katori1} and experimental
\cite{Katori2,Ye1,Bar06,Lemonde1,Ye2} breakthrough in the field of fundamental laser metrology based on a
possibility to work with strongly forbidden optical transitions, using large number of neutral atoms confined to
an optical lattice at the magic wavelength. This great advance expressed itself in a transition to spectral
structures with Hertz \cite{Bar06,Ye2} and potentially sub-Hertz widths, that  approaches a prospect of the
creation of atomic optical frequency standards with unprecedented fractional frequency uncertainty and accuracy at
a level 10$^{-17}$-10$^{-18}$.

However the achievement of such extraordinary metrological
characteristics in real frequency standards is a challenging goal.
On the way to this goal it will be necessary to determine sources
and values of systematic errors. Therefore now there comes a
period, when frequency shifts of various origin should be
thoroughly investigated. For example, the second-order (in
intensity) light shift due to the atomic hyperpolarizability has
been studied experimentally \cite{Lemonde1,Bar08} and
theoretically \cite{TY06_2}.

In the present paper, taking into account contributions due to magneto-dipole and quadrupole transitions, we
investigate a previously unknown frequency light shift caused by the quantization of atomic translational motion
in an optical lattice for a forbidden optical transition $J_g$=0$\to$$J_e$=0 (for instance, $^1S_0$$\to$$^3P_0$
in alkaline-earth-like atoms). This shift is proportional to the square root (in the Lamb-Dicke regime) from the
lattice field intensity, and it does not vanish at the magic wavelength $\lambda_m$, at which the first-order (in
intensity) light shift cancels. Preliminary estimates show that this unavoidable shift has a principal
significance for the definition of metrological characteristics of lattice-based atomic clocks.

Note, the influence of the electric quadrupole ($E2$) and magnetic
dipole ($M1$) effects on the clock levels in Sr atoms was first
discussed in \cite{Katori1}. The calculations have demonstrated
that the $E2$ and $M1$ dynamic polarizabilities at the magic
wavelength is of order $10^{-7}$ of the electric dipole ($E1$)
polarizability. However, the calculations in \cite{Katori1}
referred to the usual plane-wave approach which did not account
for the specific inhomogeneous distribution in space of electric
and magnetic field components of the optical lattice. The effects
of inhomogeneous field distribution of a standing wave are
considered in the present paper in detail.

Consider an atom confined to an optical lattice, which is induced
by a one-dimensional elliptically polarized standing wave (with
the frequency $\omega$). The electric field vector has the form:
\begin{equation}\label{E}
{\bf E}({\bf r},t)=E_0{\bf e}\cos({\bf k}\cdot{\bf r})\,e^{-i\omega t}+c.c.\,,
\end{equation}
where $E_0$ is the scalar amplitude; ${\bf k}$ is the wavevector
($k$=$|{\bf k}|$=$\omega$/$c$); ${\bf e}$ is the complex unit
polarization vector $({\bf e}\cdot{\bf e}^*)$=1. The condition
$({\bf e}\cdot{\bf k})$=0 is satisfied due to the transversality
of electromagnetic field.

First we consider the frequency shift of a transition
$J_g$=0$\to$$J_e$=0 in a potential induced only by the
contributions of electro-dipole transitions $J_j$=0$\to$$J$=1
($j$=$g$,$e$). For the standing-wave field (\ref{E}) the light
shift (potential) of a $j$-th level is spatially modulated and it
has the following form:
\begin{equation}\label{UE}
U^E_{j}({\bf r})=-{\cal W}_{j}^{}\cos^2(kz)\,;\quad {\cal W}_{j}^{}>0\quad (j=g,e)\,.
\end{equation}
Here, for convenience, we direct the $z$ axis along the wavevector
${\bf k}$. The potential amplitude ${\cal W}_{j}^{}$ depends on
the frequency $\omega$ and it is proportional to $|E_0|^2$.

Then we quantize the translational motion of atoms. To do this we
will assume the Lamb-Dicke regime, when atoms are localized in the
field antinodes  $kz$=$l$$\pi$ ($l$=0,$\pm 1$,$\pm 2$...) on the
size much less than the wavelength. In this case we can describe
the atomic motion in the harmonic approximation. For the sake of
definiteness, we will work near the point $z$=0. At the condition
$|kz|$$\ll$1 we can use the approximation
$\cos^2(kz)$$\approx$1$-$$k^2$$z^2$, which allows us to write the
shift (\ref{UE}) as the harmonic oscillator potential:
\begin{equation}\label{UE1}
U^E_{j}({\bf r})\approx -{\cal W}_{j}^{}+\frac{M(2\pi\Omega_j)^2z^2}{2}\,;\qquad (j=g,e)\,,
\end{equation}
where $M$ is the atomic mass, and the oscillator frequency
$\Omega_j$ for $j$-th level has the form:
\begin{equation}\label{Omega0}
\Omega_j=\frac{1}{2\pi}\sqrt{\frac{2{\cal W}_{j}^{}\,k^2}{M}}\,;\qquad (j=g,e)\,.
\end{equation}
The following relationships are obvious:
\begin{equation}\label{prop}
{\cal W}_{j}^{}\propto I\,,\quad \Omega_j\propto\sqrt{I}\,,
\end{equation}
where $I$=$c$$|E_0|^2$/2$\pi$ is the field intensity in the
lattice antinode.

Using for the potential (\ref{UE1}) the standard theory of
harmonic oscillator, we write the energies of the upper and lower
levels with consideration for the vibrational structure:
\begin{equation}\label{Ej}
{\cal E}_j(n)={\cal E}_j^{(0)}-{\cal W}_{j}^{}+h\Omega_j(n+1/2)\,,\quad (j=g,e),
\end{equation}
where $h$=2$\pi$$\hbar$; ${\cal E}_j^{(0)}$ is the energy of unperturbed $j$-th state in a free space;
$n$=0,1,2,... is the vibrational quantum number (see in Fig.1).

\begin{figure}[t]
\centerline{\scalebox{0.4}{\includegraphics{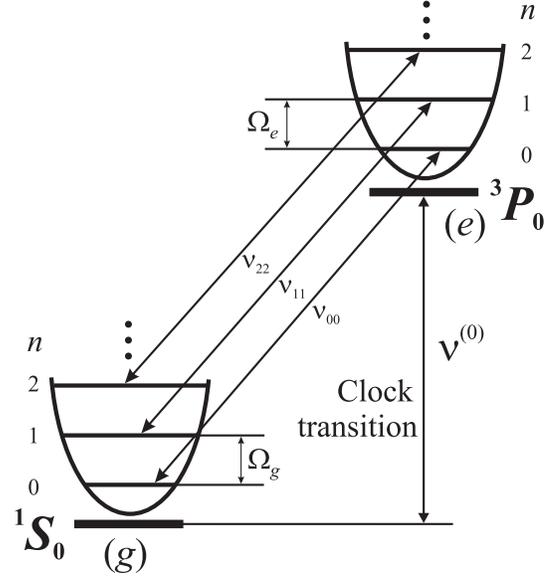}}}\caption{An
illustration of optical transitions between vibrational levels for
atoms confined to an optical lattice.}
\end{figure}

Consider the optical frequency of transition between vibrational levels with the same quantum numbers $n$:
\begin{eqnarray}\label{v_nn}
&&\nu_{nn}=\frac{{\cal E}_e(n)-{\cal E}_g(n)}{h}=\\
&&\nu^{(0)}-({\cal W}_{e}^{}-{\cal W}_{g}^{})/h+(\Omega_e-\Omega_g)(n+1/2)\,,\nonumber
\end{eqnarray}
where $\nu^{{(0)}}$=(${\cal E}_e^{(0)}$$-$${\cal E}_g^{(0)}$)/$h$
is the frequency of unperturbed $J_g$=0$\to$$J_e$=0 transition.
Taking into account the relationships (\ref{prop}), the frequency
shift can be written as:
\begin{eqnarray}\label{Dv_nn}
\Delta\nu_{nn}\equiv\nu_{nn}-\nu^{{(0)}}=\alpha(\omega)I+(n+1/2)\beta(\omega)\sqrt{I}\,,
\end{eqnarray}
where the coefficients $\alpha(\omega)$ and $\beta(\omega)$ are
individual characteristics of chosen element. They are defined as
follows:
\begin{equation}\label{ab}
 \alpha(\omega)I=-({\cal W}_{e}^{}-{\cal W}_{g}^{})/h\,;\;\;\beta(\omega)\sqrt{I}=\Omega_e-\Omega_g\,.
\end{equation}
Thus, despite of the fact that we consider the potential
(\ref{UE}), which is proportional to $I$, the effects of the
quantization of atomic motion lead (in the Lamb-Dicke regime) to
the appearance of the additional square root dependence
($\propto$$\sqrt{I}$) for the frequency shift (\ref{Dv_nn}). Note
that beyond the Lamb-Dicke regime the intensity dependence of the
frequency shift is more complicated due to the anharmonicity.

In the case of potential (\ref{UE}), which is induced by electro-dipole transitions only, at the magic frequency
$\omega_m$=2$\pi$$c$/$\lambda_m$ we have simultaneously $\alpha(\omega_m)$=0 and $\beta(\omega_m)$=0, because
${\cal W}_{e}^{}(\omega_m)$=${\cal W}_{g}^{}(\omega_m)$. However, if we take into account contributions due to
magneto-dipole and quadrupole transitions, it will not be so, i.e. $\beta(\omega_m)$$\neq$0, that can have a
principal importance for metrological characteristics. Let us prove this.

For the standing-wave field (\ref{E}) the magnetic field vector
${\bf B}$ has the form:
\begin{equation}\label{M}
{\bf B}({\bf r},t)=B^{}_0{\bf e}^{}_B\sin({\bf k}\cdot{\bf r})\,e^{-i\omega t}+c.c.\,,
\end{equation}
where $B^{}_0$=$iE^{}_0$ is the scalar amplitude of the magnetic
field, and ${\bf e}^{}_B$=$[{\bf k}$$\times$${\bf e}]$/$k$ is the
unit polarization vector of the magnetic field. As a result, the
contribution to the potential of  $j$-th level ($\propto$$|{\bf
B}|^2$) due to magneto-dipole transitions $J_j$=0$\to$$J$=1 has
the spatial dependence different from the potential (\ref{UE}):
\begin{equation}\label{UM}
U^B_{j}({\bf r})={\cal B}_{j}^{}\sin^2(kz)\,;\qquad (j=g,e)\,,
\end{equation}
where the potential amplitude ${\cal B}_{j}^{}$ is proportional to
the intensity $I$, but the frequency dependence (on $\omega$) of
${\cal B}_{j}^{}$ differs from ${\cal W}_{j}^{}$. It can be easily
shown that the contribution due to quadrupole transitions
$J_j$=0$\to$$J$=2 has the same spatial dependence (for 1D standing
wave):
\begin{equation}\label{UQ}
U^Q_{j}({\bf r})={\cal Q}_{j}^{}\sin^2(kz)\,;\qquad (j=g,e)\,.
\end{equation}
Hence, for $j$-th level the total potential proportional to the
intensity $I$ has the form:
\begin{eqnarray}\label{U_tot}
&&U^{}_{j}({\bf r})=U^E_{j}({\bf r})+U^B_{j}({\bf r})+U^Q_{j}({\bf r})=\\
&&-{\cal W}_{j}^{}\cos^2(kz)+\{{\cal B}_{j}^{}+{\cal Q}_{j}^{}\}\sin^2(kz);\quad (j=g,e).\nonumber
\end{eqnarray}
As is seen, the spatial dependencies of potentials $U^{}_{g}({\bf
r})$ and $U^{}_{e}({\bf r})$ for the lower and upper levels of the
forbidden $J_g$=0$\to$$J_e$=0 transition are different due to the
contributions induced by magneto-dipole and quadrupole
transitions, i.e. $U^{}_{g}({\bf r})$$/$$U^{}_{e}({\bf
r})$$\neq$$\,const$. As a consequence, strictly speaking, the
condition $U^{}_{e}({\bf r})$=$U^{}_{g}({\bf r})$ does not hold
for any frequency. Hence, the ideal notion of the magic frequency
$\omega_m$ does exist only for a single running wave, for which,
however, the confining optical lattice potential is absent
(spatially uniform light shift) and therefore this case is not
important for lattice-based atomic clock. However, since the
contribution due to electro-dipole transitions dominates, the
other contributions can be considered as very small perturbation,
which, nevertheless, can influence the metrological
characteristics of frequency standards.

Expanding the expression (\ref{U_tot}) in $(kz)$ powers and using
the harmonic approximation (i.e.
$\cos^2(kz)$$\approx$1$-$$k^2$$z^2$ and
$\sin^2(kz)$$\approx$$k^2$$z^2$), we obtain the expression for the
potential  analogous to (\ref{UE1}):
\begin{equation}\label{U_tot1}
U^{}_{j}({\bf r})\approx -{\cal W}_{j}^{}+\frac{M(2\pi\widetilde{\Omega}_j)^2z^2}{2}\,;\qquad (j=g,e)\,,
\end{equation}
from which all the following formulae (\ref{prop})-(\ref{ab}) are deduced. However instead of  $\Omega_j$ (see
eq.(\ref{Omega0})) now we have to use the other expression for the vibrational frequency $\widetilde{\Omega}_j$:
\begin{equation}\label{Omega_t}
\widetilde{\Omega}_j=\frac{1}{2\pi}\sqrt{\frac{2\left\{{\cal W}_{j}^{}+{\cal B}_{j}^{}+{\cal
Q}_{j}^{}\right\}k^2}{M}};\quad (j=g,e).
\end{equation}
From the equations (\ref{U_tot1}) and (\ref{Omega_t}) it follows
that for 1D standing wave the magneto-dipole and quadrupole
transitions affect only the coefficient $\beta(\omega)$ in the
formula for the shifts (\ref{Dv_nn}), while the coefficient
$\alpha(\omega)$ is governed by the electro-dipole transitions
solely as before.

The magic frequency of lattice field $\omega_m$ can now be
determined from the condition of cancelation of the linear (with
respect to the intensity $I$) shift in (\ref{Dv_nn}) (i.e.
$\alpha(\omega_m)$=0). In this case we find that ${\cal
W}_{e}^{}(\omega_m)$=${\cal W}_{g}^{}(\omega_m)$=${\cal W}$. Then
the remaining shift ($\propto$$\sqrt{I}$) in eq.(\ref{Dv_nn})
differs from zero:
\begin{equation}\label{bI}
\Delta\nu_{nn}=(n+1/2)\beta(\omega_m)\sqrt{I}\neq 0\,.
\end{equation}
Expanding the expression (\ref{Omega_t}) in small parameter
$|$$({\cal B}_{j}^{}$+${\cal Q}_{j}^{})$/${\cal W}$$|$$\ll$1 and
leaving only the first-order term, we obtain:
\begin{equation}\label{bI_1}
\beta(\omega_m)\sqrt{I}=(\widetilde{\Omega}_e-\widetilde{\Omega}_g)\approx{\Omega}^{(0)}\eta\,.
\end{equation}
Here the frequency $\Omega^{(0)}$ is equal to:
\begin{equation}\label{Omega_t0}
{\Omega}^{(0)}=\frac{1}{2\pi}\sqrt{\frac{2{\cal W}k^2}{M}}\,,
\end{equation}
and its value coincides with the vibrational frequency, which is
practically the same (at $\omega_m$) for the upper and lower
levels of the clock transition $J_g$=0$\to$$J_e$=0. The
dimensionless small coefficient  $\eta$ in (\ref{bI_1}) does not
depend on the intensity $I$ and polarization ${\bf e}$, and it is
defined as:
\begin{eqnarray}\label{eta}
&&\eta=\eta^{}_B+\eta_Q^{}\,;\\
&&\eta^{}_B=\frac{{\cal B}_{e}^{}-{\cal B}_{g}^{}}{2{\cal W}};\quad\eta^{}_Q=\frac{{\cal Q}_{e}^{}-{\cal
Q}_{g}^{}}{2{\cal W}};\quad |\eta^{}_B,\eta_Q^{}|\ll 1.\nonumber
\end{eqnarray}
The first term $\eta^{}_B$ in (\ref{eta}) is governed by the
magneto-dipole transitions, and the second term $\eta_Q^{}$ is
caused by the quadrupole transitions.

Now let us estimate the metrological significance of the
unavoidable square-root shift (\ref{bI_1}). Basing on the very
general estimations, we can expect that for different elements the
value of the coefficient $\eta$ is of order of
10$^{-7}$-10$^{-6}$. For typical optimal (with respect to the
intensity $I$) experimental conditions the vibrational frequency
${\Omega}^{(0)}$ is of order of several tens kHz. Then even for
$\eta$$\sim$10$^{-7}$ the shift (\ref{bI})-(\ref{bI_1}) is
estimated as $\Delta\nu_{nn}$$\sim\,$0.01 Hz, what is principal
for frequency standards with the accuracy 10$^{-17}$-10$^{-18}$.
It is very important that this shift can not be substantially
reduced by the decrease of the field intensity in view of weak
square-root dependence $\sqrt{I}$ (in contrast with the
hyperpolarizability leading to the shift $\propto$${I}^2$). Note
also that the discussed shift should be taken into consideration
in experiments on the precision measurement of the magic frequency
$\omega_m$.

The results obtained above can be generalized to the case of
arbitrary field configuration (including 2D and 3D optical
lattices), when the electric field vector has the general form:
\begin{equation}\label{E_gen}
{\bf E}({\bf r},t)={\bf E}({\bf r})e^{-i\omega t}+c.c.;\;\;{\bf E}({\bf r})=\sum_{a}{\bf E}_ae^{i({\bf k}_a{\bf r
})},
\end{equation}
where ${\bf E}_a$ is the vector amplitude of $a$-th running wave
with the wavevector ${\bf k}_a$ ($|$${\bf
k}_a$$|$=$k$=$\omega$/$c$). The spatial dependence of potential
induced by the electrodipole transitions is governed by the
expression:
\begin{equation}\label{UE_gen}
U^{E}_j({\bf r})=w^{}_j(\omega)\,|{\bf E}({\bf r})\,|^2\,;\qquad (j=g,e)\,,
\end{equation}
where the frequency dependence $w^{}_j(\omega)$ is an individual characteristics of chosen element.

The potential induced by the magneto-dipole transitions has the
form:
\begin{eqnarray}\label{UB_gen}
&&U^{B}_j({\bf r})=b^{}_j(\omega)\,|{\bf B}({\bf r})\,|^2\,,\qquad (j=g,e)\,;\\
&&{\bf B}({\bf r})=\sum_{a}e^{i({\bf k}_a{\bf r })}[{\bf n}^{}_{{\bf k}_a}\times{\bf E}_a]\,,\quad ({\bf
n}^{}_{{\bf k}_a}={\bf k}_a/k)\,.\nonumber
\end{eqnarray}
And, eventually, the contribution due to quadrupole transitions
can be presented in the form of the following scalar product:
\begin{equation}\label{UQ_gen}
U^{Q}_j({\bf r})=r^{}_j(\omega)\,({\bf Q}^{}_2\cdot{\bf Q}^{*}_2)\,;\qquad (j=g,e)\,,
\end{equation}
where the covariant components of irreducible tensor of the second rank ${\bf Q}^{}_{2}$ are written as:
\begin{equation}\label{Q_tenz}
{\bf Q}^{}_{2q}=\sum_{a}e^{i({\bf k}_a{\bf r })}\{{\bf n}^{}_{{\bf k}_a}\otimes{\bf E}_a\}^{}_{2q},\quad (q=0,\pm
1,\pm 2).
\end{equation}
The definition of the tensor product of two arbitrary vectors
$\{{\bf a}$$\otimes$${\bf b}\}^{}_{2q}$, and the expression for
the scalar product of the type ($\{{\bf a}$$\otimes$${\bf
b}\}^{}_{2}$$\cdot$$\{{\bf c}$$\otimes$${\bf d}\}^{}_{2}$) can be
found, for example, in the book \cite{varsh75}.

Evidently, that in the general case all the spatial dependencies (\ref{UE_gen}), (\ref{UB_gen}) and
(\ref{UQ_gen}) are different from each other. In this case the magneto-dipole and quadrupole transitions can also
affect the linear term in (\ref{Dv_nn}), i.e. the coefficient  $\alpha(\omega)$. This will take place, if in
minima points $\{{\bf r}_{min}\}$ of the electro-dipole potential $U^{E}_j({\bf r})$ we have $U^{B}_j({\bf
r}_{min})$$\neq\,$0 and/or $U^{Q}_j({\bf r}_{min})$$\neq\,$0 (as distinct from the ideal one-dimensional standing
wave (\ref{E})). As the result the magic frequency  $\omega_m$ will differ from its value for the 1D standing
wave, for the later it is worth to introduce a separate notation $\omega^{E}_m$. In the order of magnitude the
region of possible variations of $\omega_m$ for different field configurations can be roughly estimated as
$\sim\,$$\eta\,$$\omega^{E}_m$, i.e. for some elements one can expect the variation at a level 100 MHz (or even
more). Even in the one-dimensional case, when the counterpropagating waves are unbalanced (i.e. when they have
different amplitudes) the magneto-dipole and quadrupole contributions have influence on the coefficient
$\alpha(\omega)$ in (\ref{Dv_nn}), changing the magic frequency $\omega_m$.

Concluding, in the present paper for a strongly forbidden optical
transition $J$=0$\to$$J$=0, taking into account contributions due
to the magneto-dipole and quadrupole transitions, we have
investigated the previously unknown frequency light shift caused
by the quantization of atomic translational motion in an optical
lattice. Using 1D standing wave as an example, we have proved that
the above mentioned factors leads to the existence of the
additional frequency shift, which has square-root dependence on
the lattice field intensity in the Lamb-Dicke regime. It has been
also shown that this shift does not vanish at the magic
wavelength. The main intrigue is that the new square-root
frequency shift can have a principal significance for the
metrological characteristics of atomic clocks and, due to this
reason, it should be thoroughly investigated. In view of the weak
square-root dependence $\propto$$\sqrt{I}$ and small absolute
value, it maybe difficult to measure experimentally this shift
with a good accuracy (especially against the background of strong
dependence $\propto$$\,I^2$ due to the hyperpolarizability), what
increases the importance of theoretical calculations. Apart from
this, it has been found that the numerical value of the magic
frequency depends on the concrete configuration of the lattice
field and it can vary within the limits 1-100 MHz (depending on
element) as one passes from one field configuration to another.

The obtained results can be significant for the choice of
experimental methods in the case, when the unavoidable shift turns
out to be important. For example, from eq.(\ref{bI}) it follows
that the shift $\Delta\nu_{nn}$ depends on the vibrational quantum
number $n$. Therefore, before the beginning of spectroscopic
measurements it is advisable to transfer atoms to the lowest
vibrational level with  $n$=0 (similarly to the method
successfully realized in \cite{Lemonde3}). Besides, it is
preferable to use a resonator for the formation of 1D standing
wave for the purposes of more reliable balance of
counterpropagating waves.

A.V.T. and V.I.Yu. were supported by RFBR (07-02-01230,
07-02-01028, 08-02-01108), INTAS-SBRAS (06-1000013-9427) and
Presidium of SB RAS. V.D.O. was supported by RFBR (07-02-00279),
CRDF and MinES RF (ANNEX-BP2M10).

\end{document}